\documentclass[a4paper]{article}

\usepackage{amsmath}
\usepackage{amssymb}
\usepackage{bbm}
\usepackage[dvips]{graphicx}

\newcommand{\diff}{\mathrm{d}}

\begin{document}

\begin{center}
\bf \LARGE An effective model for flocculating bacteria \\
with density-dependent growth dynamics
\end{center}

\bigskip

\begin{center}
B.~Haegeman$^{1,2}$, C.~Lobry$^{1}$, J.~Harmand$^{1,2}$
\end{center}

\bigskip

{\small \begin{enumerate}
\item MERE INRIA--INRA Research Team, UMR ``Analyse des Syst\`emes et Biom\'etrie'', INRA, 2 Place Pierre Viala, 34060 Montpellier, France.
\item Laboratoire de Biotechnologie de l'Environnement, INRA--LBE, Avenue des \'Etangs, 11100 Narbonne, France.
\end{enumerate} }

\bigskip

\begin{quote}
We present a model for a biological reactor in which bacteria tend to aggregate in flocs, as encountered in wastewater treatment plants. The influence of this flocculation on the growth dynamics of the bacteria is studied. We argue that a description in terms of a specific growth rate is possible when the flocculation dynamics is much faster than the other processes in the system. An analytical computation shows that in this case, the growth rate is density-dependent, i.e., depends both on the substrate and the biomass density. When the flocculation time scale overlaps with the other time scales present in the system, the notion of specific growth rate becomes problematic. However, we show numerically that a
density-dependent growth rate can still accurately describe the system response to certain perturbations.
\end{quote}

\bigskip

\section{Introduction}

Biological reactors are commonly used to remove pollutants from wastewaters. One standard technology is the two-step Activated Sludge Process (ASP). Both in the reaction and the settling tank, bacteria naturally aggregate and form flocs. It is well known -- but poorly understood -- that the flocculation is strongly dependent on the mass and volumetric loading rates: the higher the loading rate, the higher the risk of decreasing sludge floc formation and settling capacity. In order to optimize this bioprocess, it is therefore important to better understand the flocculation phenomenon.

Mathematical modelling has turned out to be a valuable tool in the study of WasteWater Treatment Plants (WWTP). To coordinate efforts in the development of efficient design and control tools, the Activated Sludge Model No.~1 (ASM1) was proposed in the late eighties. This model describes the different biological processes (e.g., chemical oxygen demand (COD) removal, (de)nitrification and phosporous removal) in detail. Its core consists of the mass balance equations, including the reaction kinetics as a function of the limiting substrates, which read in their simplest form,
\begin{align}
 \frac{\diff x}{\diff t} &= h(s) x - D x \nonumber \\
 \frac{\diff s}{\diff t} &= -h(s) x + D (s_\text{in}-s),
 \label{eq:homokinetics}
\end{align}
where $x$ is the biomass concentration, $s$ the substrate concentration, $h(s)$ the specific growth rate, $D$ the dilution rate and $s_\text{in}$ the substrate concentration in the inflow.

Although a number of improvements have been investigated for the reaction dynamics (e.g., ASM2, ASM2d and ASM3), the model part for the floc formation and settling remains the weakest part. This modelling problem (see \cite{E97, TJF99} for reviews) has been studied from different perspectives. Population Balance Models (PBM) describe the floc aggregation and breakage and allow to compute the floc size distribution as a function of time \cite{BL02,N05,DHB06}. Computational Fluid Dynamics (CFD) simulators describe the hydrodynamics in the clarification tank and try to predict the settling properties of the flocs \cite{AKR00}. Individual-Based Models (IBM) take both physico-chemical and biological processes into account at the level of a single floc \cite{M04}.

These modelling approaches have in common a high-dimensional parameter space. Although these parameters can be identified from experiments, the resulting model is often too complex to provide insight in the governing mechanisms. Moreover, to compute, for instance, the settling properties of the ensemble of interacting flocs in the clarifier, one has to combine a CFD with a PBM approach, which leads to even more intricate models.

Instead of starting from performant simulators, we propose to take the simple model (\ref{eq:homokinetics}) as a point of reference. In particular, we investigate how these equations are modified when the biomass is organized in flocs. To address this question, we propose a PBM-like model where both the floc interactions (as in standard PBM) and the bacteria growth are included. This qualitative model is sufficiently transparent to be manipulated analytically. Our approach is primarily intended to model the floc dynamics in the reaction tank, where both physico-chemical and biological processes have to be taken into account. Nevertheless, our model can also be useful to check the common assumption of PBM that biological growth can be neglected in the settling tank.

Our analysis naturally leads to an effective model of the form,
\begin{align}
 \frac{\diff x}{\diff t} &= h(s,x) x - D x \nonumber \\
 \frac{\diff s}{\diff t} &= -h(s,x) x + D (s_\text{in}-s).
 \label{eq:heterokinetics}
\end{align}
Note that the specific growth rate $h(s,x)$ depends both on the substrate concentration $s$ and the biomass concentration $x$, in contrast with the substrate-dependent growth rate $h(s)$ of model (\ref{eq:homokinetics}). The specific growth rate $h(s,x)$ is called density-dependent. In fact, based on the original work by Arditi and Ginzburg \cite{AG89}, density-dependent growth rates were recently proposed to describe bioreactor kinetics more accurately \cite{LH06}. From an ecological point of view, this change has important consequences, as it allows microorganisms to coexist in a medium where classical, i.e. substrate-dependent, models predict extinction by wash-out.

It should be noted that this work is not the first to study the influence of a heterogeneous biomass structure on the growth rate (see, for example, \cite{PMA98,C99}). However, we present here, to the best of our knowledge, an original derivation of an effective model with density-dependent growth dynamics, starting from a PBM description including bacterial growth. 

This paper is organized as follows. In Section 2, we introduce the bioreactor model. The different phenomena, including bacterial growth, floc aggregation and breakage, and hydrodynamics, are discussed. In Section 3, we perform an analytical study of the model, under the hypothesis that the time scale associated with the floc interactions is much shorter than the other processes present in the system. We show analytically how this hypothesis leads to a density-dependent growth rate. In Section 4, we present some numerical computations, that go beyond the hypothesis of separate time scales.

\section{Modelling flocculation of growing bacteria}

We start by introducing some notation. Consider a bioreactor in which a biomass grows on a substrate. The density of the biomass is denoted by $x$, the density of the substrate by $s$. The biomass consists of bacteria which naturally aggregate in flocs. A floc containing $n$ bacteria will be denoted by $F_n$. Define $u_n$ as the density of flocs of size $n$. Expressing the densities $x$ resp.\ $u_n$ as the number of particles (bacteria resp.\ flocs) per unit of volume, we have
\begin{equation}
 x = \sum_{n=1}^\infty n u_n. \label{eq:bactdensity}
\end{equation}

We assume the reactor to be perfectly mixed. As a consequence, all flocs have the same access to the substrate. However, we will take into account that the bacteria at the surface of the flocs have easier access to the substrate than the bacteria inside the flocs.

The dynamics of the floc densities $u_n$ is given by
\begin{equation}
 \frac{\diff u_n}{\diff t}
 = \left( \frac{\diff u_n}{\diff t} \right)_\text{bacterial growth}
 - D\,u_n
 + \left( \frac{\diff u_n}{\diff t} \right)_\text{floc interaction}.
 \label{eq:model}
\end{equation}
The second term in the right-hand side represents the bacteria disappearing in the effluent of the reactor with dilution rate $D$.
The other two terms, due to the bacterial growth and the floc interaction, are now described in more detail.

\subsection*{Dynamics of bacterial growth}

The only bacterial growth present in our model is through cell division. As a bacterium present in a floc of size $n$ divides, we assume the daughter bacteria to stick to the floc, which will then consists of $n+1$ bacteria. This growth can be written as
\[
 F_n \to F_{n+1}
 \qquad \text{with reaction rate $h_n(s)$.}
\]
Note that the growth rate of a floc depends on its size $n$ and on the substrate density $s$. To describe that the substrate has immediate access to the outer shell of the floc, whereas the inner bacteria can be deprived from the substrate, we assume the dependency on $n$ and $s$ to be
\begin{equation} \label{eq:growthfractal}
 h_n(s) = h(s) n^\alpha,
\end{equation}
with the exponent $\alpha\leq 1$. When the access to the substrate is not limited, $\alpha=1$; when it is, $\alpha<1$ and its value corresponds to the surface-to-volume ratio of the flocs. For spherical flocs, $\alpha=\frac 23$. As flocs are known to have some fractal form, also other exponents are possible.

As will become clear from the computations, our results do not critically depend on the function $n \mapsto h_n(s)$. A more complicated sublinear dependence than (\ref{eq:growthfractal}) can easily be handled, taking the hydrodynamics around the floc into account \cite{H95}, or using the results of an individual-based model \cite{M04}. As our treatment is qualitative, and the precise behaviour of $h_n(s)$ therefore of secondary importance, we will restrict our attention to floc growth functions of the form (\ref{eq:growthfractal}).

The dynamics corresponding to bacterial growth is
\begin{align}
 \left( \frac{\diff u_1}{\diff t} \right)_\text{bacterial growth}
 &= - h_1(s) u_1 \nonumber \\
 \left( \frac{\diff u_n}{\diff t} \right)_\text{bacterial growth}
 &= h_{n-1}(s) u_{n-1} - h_n(s) u_n,
 \qquad n\geq 2. \label{eq:growthdyna}
\end{align}
Indeed, a growth event $F_n \to F_{n+1}$ corresponds to the consumption of a floc of size $n$ and the production of a floc of size $n+1$. Mass action kinetics are assumed for this reaction.

\subsection*{Aggregation--breakage dynamics}

The floc interactions we consider are the aggregation of two flocs to form one bigger floc and the breakage of one floc into two smaller ones. As equations (\ref{eq:model}) are continuous in time, processes involving three or more flocs are implicitly included. The floc interactions can be written as
\begin{align}
 & F_m + F_n \to F_{m+n} \qquad
   \text{with reaction rate $a_{m,n}$,} \nonumber \\
 & F_{m+n} \to F_m + F_n \qquad
   \text{with reaction rate $b_{m,n}$.} \label{eq:flocreactgeneral}
\end{align}
The reaction rates are symmetric in their arguments, i.e., $a_{m,n}=a_{n,m}$ and $b_{m,n}=b_{n,m}$. Many studies have been carried out to derive theoretical relations between these coefficients \cite{S17,TJF99}, or to identify them from experiments \cite{BL02,N02,DHB06}. We note already that our analysis, in the first place qualitative, will not need explicit expressions for the reaction rates $a_{m,n}$ and $b_{m,n}$.

The part of the dynamics corresponding to the floc interactions is
\begin{align}
 \left( \frac{\diff u_n}{\diff t} \right)_\text{floc interaction}
 =& \sum_{m=1}^{\lfloor \frac n2 \rfloor} a_{m,n-m} u_m u_{n-m}
 - \sum_{m=1}^{\infty} (1+\delta_{m,n)} a_{m,n} u_m u_n \nonumber \\
 &+ \sum_{m=1}^{\infty} (1+\delta_{m,n}) b_{m,n} u_{m+n}
 - \sum_{m=1}^{\lfloor \frac n2 \rfloor} b_{m,n-m} u_n, \label{eq:flocdynageneral}
\end{align}
where $\lfloor x \rfloor$ denotes the largest integer smaller than
$x$, and
\[
 \delta_{m,n}
 = \begin{cases}
   1 & \text{if $m=n$,} \\
   0 & \text{otherwise.}
   \end{cases}
\]
The first term corresponds to the aggregation of two flocs to form a floc $F_n$. The second term corresponds to the aggregation of a floc $F_n$ with another floc. The third term corresponds to the breakage of a floc into two flocs, one of which has size $n$. The fourth term corresponds to the breakage of a floc $F_n$ into two smaller ones.

The equations (\ref{eq:flocdynageneral}) are identical to the so-called population balance models (PBM), first introduced by Smoluchowski \cite{S16} and extensively used in flocculation modelling \cite{HRM88,TJF99,R00,BL02,N02,DHB06}. As also the bacterial growth (\ref{eq:growthdyna}) is compatible with the population balance structure, our model can be considered as an extension of PBM.

In contrast to the bacterial growth dynamics (\ref{eq:growthdyna}), the floc interactions (\ref{eq:flocdynageneral}) satisfy conservation of bacteria density, i.e.,
\[
 \sum_{n=1}^\infty n \left( \frac{\diff u_n}{\diff t} \right)
 _\text{floc interaction} = 0.
\]

\subsection*{Attachment--detachment dynamics}

For a subset of the floc interactions (\ref{eq:flocreactgeneral}), we are able to push the mathematical analysis further. In particular, we will restrict the interactions to the aggregation of a floc with a single bacterium, and the splitting off of a single bacterium from a floc. We call these processes attachment and detachment, respectively. These interactions can be written as
\begin{align}
 & F_n + F_1 \to F_{n+1} \qquad
   \text{with reaction rate $a_n$,} \nonumber \\
 & F_{n+1} \to F_n + F_1 \qquad
   \text{with reaction rate $b_n$.} \label{eq:flocreactexample}
\end{align}

The part of the dynamics corresponding to this type of floc interaction is
\begin{align}
 \left( \frac{\diff u_1}{\diff t} \right)_\text{floc interaction}
 &= - 2 a_1 u_1^2 - \sum_{n=2}^\infty a_n u_1 u_n
    + 2 b_1 u_2 + \sum_{n=3}^\infty b_{n-1} u_n \nonumber\\
 \left( \frac{\diff u_n}{\diff t} \right)_\text{floc interaction}
 &=   a_{n-1} u_1 u_{n-1} - a_n u_1 u_n
    + b_n u_{n+1} - b_{n-1} u_n,
    \qquad n\geq 2. \label{eq:flocdynaexample}
\end{align}
Since these equations are a special case of (\ref{eq:flocdynageneral}), they satisfy again conservation of bacteria density.

\section{Fast flocculation dynamics} \label{sect:withsep}

The previous section introduced the model (\ref{eq:model}), with (\ref{eq:growthdyna}) and (\ref{eq:flocdynageneral}) or (\ref{eq:flocdynaexample}). In this section we present an analytical model analysis, by assuming that the flocculation dynamics are much faster than the bacterial growth and the reactor dilution. This assumption will be relaxed in Sect.~\ref{sect:withoutsep}.

\subsection*{Separation of time scales}

Population balance models when used to describe flocculating bacteria, assume that the flocculation can be uncoupled from the other processes. It is argued that in the settling tank the substrate concentration $s$ is sufficiently low to justify this assumption. The situation is however less clear in the reaction tank. Literature reports flocculation times of the order of 1 to 10 minutes \cite{W94,DHB06}, to be compared with bacterial growth times, i.e. the inverse of the specific growth rate $h(s)$, of 1 hour to 1 day and with retention times, i.e. the inverse of the dilution rate $D$, of a few hours to a few days.

Even if the separation of time scales is not always satisfied in reality, it is interesting to investigate how our model behaves as this separation becomes infinitely large. Indeed, the model simplifies drastically and can be studied quite explicitly, as we will now show. Moreover, the reduced model can be considered as an approximation for the full model, as we will show in the next section.

To make the separation in time scales explicit, we introduce a small parameter $\epsilon >0$,
\begin{equation}
 \frac{\diff u_n}{\diff t}
 = \left( \frac{\diff u_n}{\diff t} \right)_\text{bacterial growth}
 - D\,u_n  + \frac 1\epsilon
   \left( \frac{\diff u_n}{\diff t} \right)_\text{floc interaction}.
 \label{eq:timescales}
\end{equation}
Taking $\epsilon\to 0$, we introduce a sharp distinction between
\begin{itemize}
\item the fast dynamics, consisting of the floc interaction, for times $t\sim\epsilon$, and
\item the slow dynamics, consisting of the bacterial growth and the dilution, for times $t\sim 1$.
\end{itemize}

The idea now is as follows. On the short time scale, the system evolves to fast dynamics equilibria $(u_n^\text{fast})$ which are parametrized by the total bacteria density $x$. On the large time scale, the system evolves on the manifold of these equilibrium distributions. As this manifold is one-dimensional and parametrized by $x$, we obtain autonomous dynamics for the biomass density $x$.

To be more explicit, let us write down the dynamics for $x$ by introducing (\ref{eq:bactdensity}) into (\ref{eq:model}). As the floc interactions conserve $x$,
\[
 \frac{\diff x}{\diff t}
 = \sum_{n=1}^\infty n
   \left( \frac{\diff u_n}{\diff t} \right)_\text{bacterial growth}
 - D \sum_{n=1}^\infty n u_n.
\]
Using (\ref{eq:growthdyna}) and (\ref{eq:bactdensity}),
\[
 \frac{\diff x}{\diff t}
 = \sum_{n=1}^\infty h_n(s) u_n - D\,x.
\]
At this point we use the separation of time scales. As we are looking at the slow dynamics, the distribution $(u_n)$ will have reached its fast dynamics equilibrium $\big(u_n^\text{fast}(x)\big)$. Therefore,
\begin{equation} \label{eq:reducedyna}
 \frac{\diff x}{\diff t}
 = \sum_{n=1}^\infty h_n(s) u_n^\text{fast}(x) - D\,x
 = h(s,x)\,x - D\,x,
\end{equation}
where we have introduced the specific growth rate $h(s,x)$,
\begin{equation} \label{eq:growthrate}
 h(s,x) = \frac{\displaystyle
  \sum_{n=1}^\infty h_n(s) u_n^\text{fast}(x)}{x}.
\end{equation}
Note that $h(s,x)$ depends both on the substrate concentration $s$ and the biomass concentration $x$. The obtained specific growth rate is therefore density-dependent, as announced in the introduction.

\subsection*{Uniqueness of fast dynamics equilibrium}

In the previous computation, we used the hypothesis that the fast dynamics reach the equilibrium distribution $(u_n^\text{fast}(x))$. This hypothesis can be justified by exploiting the analogy between molecules and chemical reactions on one hand, and flocs $F_n$ and floc interactions (\ref{eq:flocreactgeneral}) on the other. Indeed, the condition for chemical equilibrium can be translated to
\[
 K_{m,n} = \frac{u_{m+n}}{u_m u_n},
 \qquad
 \text{for all $m,n$},
\]
with $K_{m,n}$ the equilibrium constant, independent of any density $u_k$. Obviously, not all these conditions are independent. For example,
\[
 \begin{array}{rcl}
  F_{m} + F_{n} & \rightleftharpoons & F_{m+n} \\
  F_{m+n} + F_{1} & \rightleftharpoons & F_{m+n+1} \\
  F_{m+1} & \rightleftharpoons & F_{m} + F_{1} \\
 \hline
  F_{m+1} + F_{n} & \rightleftharpoons & F_{m+n+1},
 \end{array}
\]
and the equivalent computation for the equilibrium constants,
\[
 K_{m+1,n}
 = \frac{u_{m+n+1}}{u_{m+1} u_n}
 = \frac{u_{m+n}}{u_{m} u_n}
   \frac{u_{m+n+1}}{u_{m+n} u_1}
   \frac{u_m u_1}{u_{m+1}}
 = K_{m,n} K_{m+n,1} K_{m,1}^{-1}.
\]
To compute the fast dynamics equilibrium, it suffices to consider a basis of chemical reactions, i.e., a set of independent reactions from which the other reactions can be obtained by taking linear combinations. For our case, one such basis is given by the attachment--detachment interactions (\ref{eq:flocreactexample}). An even simpler basis is
\[
 n F_1 \rightleftharpoons F_n
 \qquad
 \text{with equilibrium constant $K_{n}$}.
\]
The equilibrium conditions then read
\begin{equation} \label{eq:fastequiproof}
 u_n = K_n u_1^n,
 \qquad
 \text{for all $n\geq 2$}.
\end{equation}
We now claim that for any total bacteria density $x$, there is only one distribution $(u_n)$ satisfying (\ref{eq:fastequiproof}) and (\ref{eq:bactdensity}). Indeed, by (\ref{eq:fastequiproof}) all $u_n$ with $n\geq 2$ are expressed in terms of $u_1$. Introducing this in (\ref{eq:bactdensity}),
\[
 x = \sum_{n=1}^\infty n u_n
   = \sum_{n=1}^\infty n K_n u_1^n.
\]
The right-hand side is a polynomial in $u_1$ with positive coefficients, and vanishes for $u_1=0$. As a consequence, for every positive $x$, there is a single $u_1$ for which this equation is satisfied. Using (\ref{eq:fastequiproof}) we obtain $u_n$ for all $n\geq 2$. This yields the unique distribution $(u_n^\text{fast}(x))$.

\subsection*{Stability of fast dynamics equilibrium}

The previous analysis started from an equilibrium description for the floc interactions. The validity of such a description can be argued by noting that on the short time scale, our model reduces to a closed system of chemical reactions. Thermodynamics guarantees that such a system evolves towards a unique equilibrium, which can be computed in terms of the equilibrium constants $K_{m,n}$ \cite{G68,V99}.

Nevertheless, it would be more satisfying to derive our conclusions directly from the kinetics (\ref{eq:flocdynageneral}) for the floc interactions. However, this is only possible under additional assumptions on the floc reaction rates $a_{m,n}$ and $b_{m,n}$. A sufficient condition is the so-called detailed balance condition, which essentially boils down to requiring the positivity of the entropy production \cite{G68,V99}. We do not develop this argument further.

It is interesting to note that in the case of attachment--detachment interactions (\ref{eq:flocreactexample}), the uniqueness and the stability of the fast dynamics equilibrium can be shown explicitly. This is consistent with our previous remark, as the detailed balance condition can be shown to become empty for these interactions. In other words, the application of the separation of time scales can be justified mathematically in the case of attachment--detachment interactions (\ref{eq:flocreactexample}).

First, we prove the uniqueness of the fast dynamics equilibrium for attach\-ment--detach\-ment. We have to find a distribution $(u_n)$ such that the right-hand side of (\ref{eq:flocdynaexample}) vanishes. As these equations satisfy conservation of bacteria density, we can neglect one of the equations, say $n=1$. Summing all the other equations, we obtain
\[
 a_1 u_1 u_1 - b_1 u_2 = 0
 \qquad \text{or} \qquad
 u_2 = \frac{a_1}{b_1} u_1^2.
\]
This relationship allows us to express all the $u_n$, $n\geq 2$ in terms of $u_1$,
\[
 u_n = \left( \prod_{k=1}^{n-1} \frac{a_k}{b_k} \right) u_1^n.
\]
As this equation has the form of (\ref{eq:fastequiproof}), the argument following the latter equation can now be repeated.

Next, to prove the stability of the fast dynamics equilibrium $(u_n^\text{fast}(x))$ for attachment--detachment, we make the additional assumption that the coefficients $a_n$ and $b_n$ have the same size dependence, i.e.,
\[
 \text{there exists a function $n \mapsto f_n$ such that}
 \ a_n = a f_n \ \text{and} \ b_n = b f_n.
\]
Consider then the entropy-like functional
\[
 H((u_n(t))) = - \sum_{n=1}^\infty u_n(t) \ln u_n(t)
  + \left( 1-\ln\frac{a}{b}\right) \sum_{n=1}^\infty u_n(t).
\]
Computing its time derivative under the fast dynamics (\ref{eq:flocdynaexample}),
\begin{align*}
 \frac{\diff H}{\diff t}
  &= - \sum_{n=1}^\infty \ln u_n \frac{\diff u_n}{\diff t}
     - \ln\frac{a}{b} \sum_{n=1}^\infty \frac{\diff u_n}{\diff t} \\
  &= \sum_{n=1}^\infty
     b u_{n+1} \ln \frac{b u_{n+1}}{a u_1 u_n} +
     a u_1 u_n \ln \frac{a u_1 u_n}{b u_{n+1}} \\
  &= \sum_{n=1}^\infty
     \left( b u_{n+1} - a u_1 u_n \right)
     \left( \ln b u_{n+1} - \ln a u_1 u_n \right) \geq 0.
\end{align*}
Moreover, it vanishes if and only if
\[
 u_{n+1} = \frac{a u_1}{b} u_n,
 \qquad \text{for all $n$}.
\]
Together with the conservation of total bacteria density $x$, we conclude that $H(t)$ is strictly increasing in time until it reaches its maximum at $(u_n)=(u_n^\text{fast}(x))$. This establishes the announced stability property.

\subsection*{Density-dependence of growth rate}

Under the assumption of fast flocculation, we derived the specific growth rate (\ref{eq:growthrate}). Together with (\ref{eq:growthfractal}) and (\ref{eq:fastequiproof}),
\[
 h(s,x) = h(s) \frac{\displaystyle \sum_n n^\alpha
 u_n^\text{fast}(x)}{x} \qquad \text{with }
 u_n^\text{fast}(x) = K_n \big(u_1^\text{fast}(x)\big)^n.
\]
For $\alpha=1$, we find $h(s,x)=h(s)$. Indeed, the case $\alpha=1$ corresponds to the assumption that all the bacteria in the flocs have the same substrate density $s$ available. The specific growth rate is then substrate-dependent. On the other hand, when $\alpha<1$, we obtain a genuine density-dependent growth rate.

The difference between substrate-dependence and density-dependence has important ecological consequences. Indeed, consider different bacterial species growing on a single substrate. A model with substrate-dependent growth rates predicts an equilibrium where only one species survives, whereas a model with density-dependent growth rates allows for an equilibrium where different species coexist. To establish the latter property, Ref.~\cite{LMR05} requires the map $x \mapsto h(s,x)$ to be decreasing. We now present a proof of this property for our growth rate $h(s,x)$.

To simplify notation, we drop the superscript ``fast''.
As $x\mapsto u_1(x)$ is increasing, it suffices to prove that
$h(s,x(u_1)) \geq h(s,x(u'_1))$ if $u_1 < u'_1$, or
\[
 \frac{\displaystyle \sum_n n^\alpha K_n u_1^n}
  {\displaystyle \sum_n n K_n u_1^n}
 \geq
 \frac{\displaystyle \sum_n n^\alpha K_n {u'_1}^n}
  {\displaystyle \sum_n n K_n {u'_1}^n}.
\]
Equivalently,
\[
 \sum_{m,n} m^\alpha K_m u_1^m n K_n {u'_1}^n
 - \sum_{m,n} m^\alpha K_m {u'_1}^m n K_n u_1^n
 \geq 0.
\]
As the terms $m=n$ cancel out, this difference can be written as
\[
 \sum_{m<n} m^\alpha n^\alpha K_m K_n u_1^m {u'_1}^m
 \big( n^{1-\alpha} - m^{1-\alpha} \big)
 \big( {u'_1}^{n-m} - u_1^{n-m} \big)
 \geq 0.
\]
Both factors in brackets are positive because $\alpha < 1$, $m <
n$ and $u_1 < u'_1$. This finishes the proof. Note also that the
inequality is strict as soon as $K_m > 0, K_n > 0, m \neq n$.

\section{Slow flocculation dynamics} \label{sect:withoutsep}

The analysis in the previous section is based on the hypothesis that the parameter $\epsilon$ in Eq.~(\ref{eq:timescales}) is small. This separation of the flocculation and the bacterial growth time scales seems to be well satisfied in the settling tank, but rather inadequate for modelling the reaction tank. In
this section we investigate how our analysis can be extended without making this assumption.

Recall the two dynamical systems we are considering. First, there are the full dynamics in terms of the densities $u_n$ of flocs of size $n$. Together with the substrate dynamics, the system is given by
\begin{align}
 \frac{\diff u_1}{\diff t}
 &= - h_1(s) u_1 - D\,u_1  + \frac 1\epsilon
    \left( \frac{\diff u_1}{\diff t} \right)_
\text{floc interaction}
 \nonumber \\
 \frac{\diff u_n}{\diff t}
 &= h_{n-1}(s) u_{n-1} - h_n(s) u_n - D\,u_n + \frac 1\epsilon
    \left( \frac{\diff u_n}{\diff t} \right)_\text{floc interaction}
 \qquad n=2,3,\ldots \nonumber \\
 \frac{\diff s}{\diff t}
 &= - \sum_{n=1}^\infty h_n(s) u_n
    + D\,(s_\text{in}-s). \label{eq:fulldyn}
\end{align}
Under the hypothesis of separate time scales, the full dynamical equations can be approximated by the reduced dynamics
\begin{align}
 \frac{\diff x}{\diff t} &= h(s,x) x - D x \nonumber \\
 \frac{\diff s}{\diff t} &= -h(s,x) x + D (s_\text{in}-s).
 \label{eq:reddyn}
\end{align}

We integrate both dynamical systems with parameter values given in Tab.~\ref{tab:compfullred}. The floc growth rate behaves as $h_n(s) \propto n^\alpha$ with $\alpha=\frac 23$, corresponding to the surface-to-volume ratio of spherical flocs. We consider attachment-detachment interactions (\ref{eq:flocdynaexample}), with reaction rates $a_n \propto n^\alpha$ and $b_n \propto n^\alpha$ with $\alpha=\frac 23$. The infinite sequence $n=2,3,\ldots$ of dynamical equations (\ref{eq:fulldyn}) is truncated at $n=300$. The parameters of Tab.~\ref{tab:compfullred} and the initial conditions in the simulations are chosen such that this truncation yields a good approximation of the full dynamics.

\begin{figure}
\begin{center}
\raisebox{4.5cm}{(a)} \hspace{-.3cm}
\includegraphics[width=5.5cm]{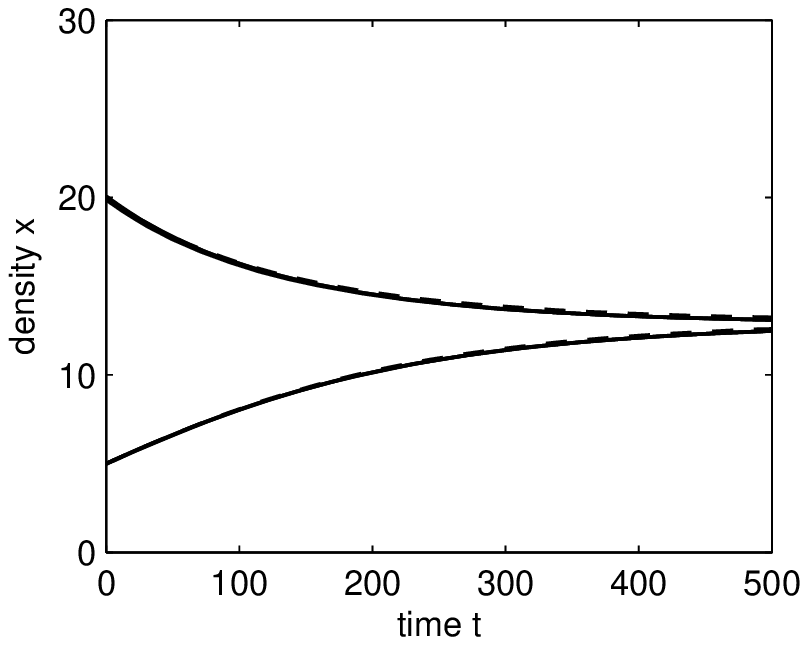}
\hspace{.5cm}
\raisebox{4.5cm}{(b)} \hspace{-.3cm}
\includegraphics[width=5.5cm]{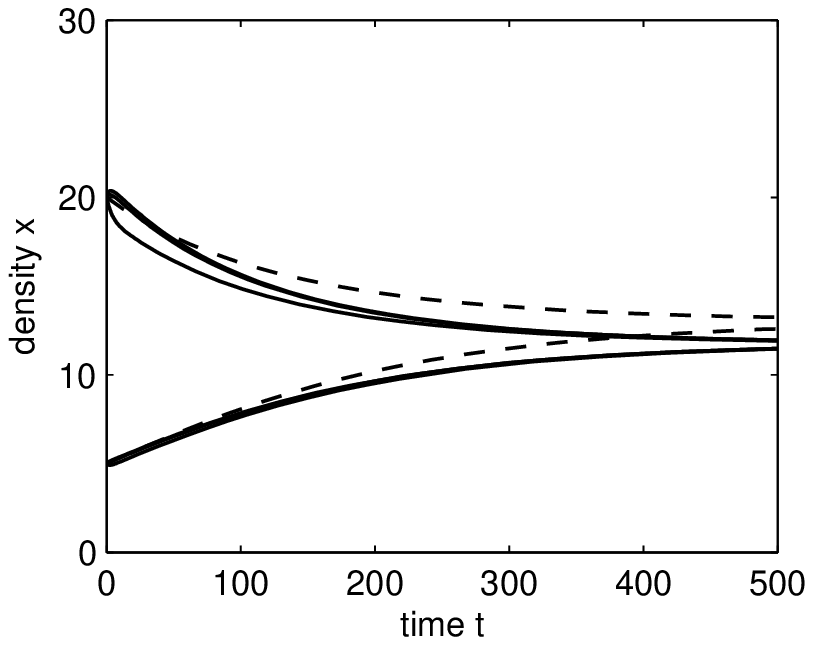} \\
\raisebox{4.5cm}{(c)} \hspace{-.3cm}
\includegraphics[width=5.5cm]{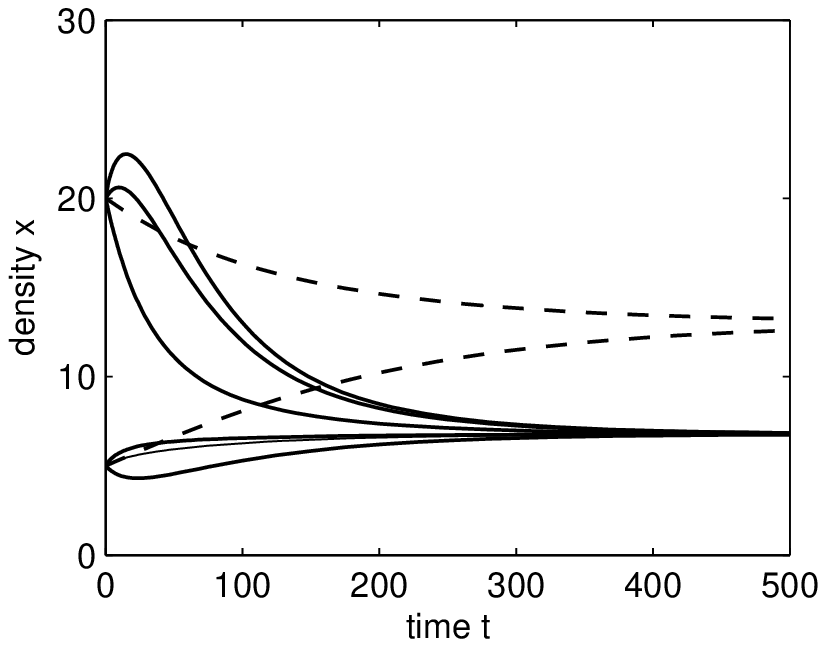}
\hspace{.5cm}
\raisebox{4.5cm}{(d)} \hspace{-.3cm}
\includegraphics[width=5.5cm]{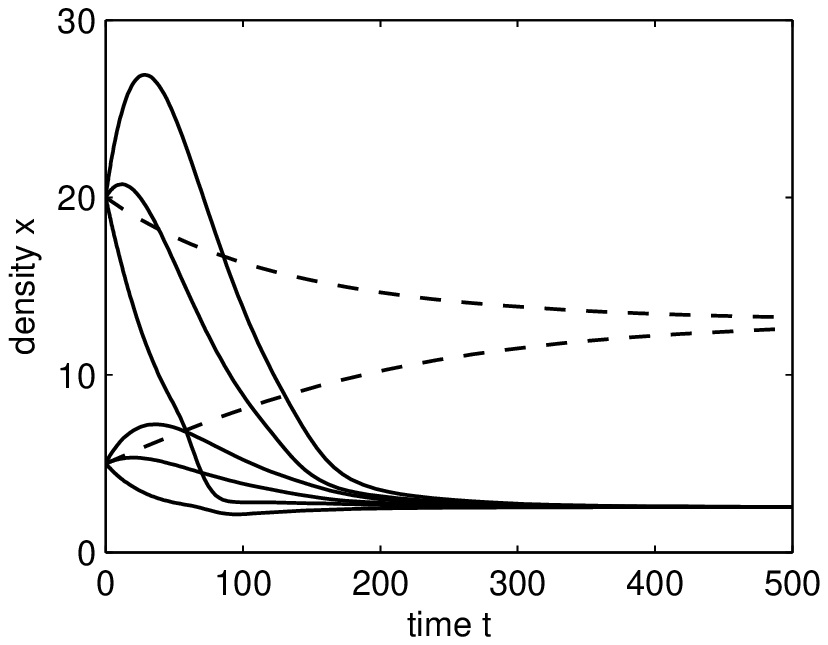}
\caption{Comparison between the full dynamics (\ref{eq:fulldyn}) for different $\epsilon$ and the reduced dynamics (\ref{eq:reddyn}). Parameter values of Tab.~\ref{tab:compfullred} were used. The full dynamics, shown in full line, were integrated for six initial conditions: three with $x(0)=20$ ($u_n(0)=20\, \delta_{n,1}$, $u_n(0)=2\, \delta_{n,10}$ and $u_n(0)=0.2\, \delta_{n,100}$) and three with $x(0)=5$ ($u_n(0)=5\, \delta_{n,1}$, $u_n(0)=0.5\, \delta_{n,10}$ and $u_n(0)=0.05\, \delta_{n,100}$). The reduced dynamics, shown in dashed line, were integrated with initial conditions $x(0)=20$ and $x(0)=5$. (a) $\epsilon=0.001$; (b) $\epsilon=0.01$; (c) $\epsilon=0.1$; (d) $\epsilon=1$.}
\label{fig:compfullred}
\end{center}
\end{figure}

Fig.~\ref{fig:compfullred} compares the full dynamics (\ref{eq:fulldyn}) for different values of the parameter $\epsilon$ with the reduced dynamics (\ref{eq:reddyn}). For small $\epsilon$, the solutions of (\ref{eq:fulldyn})
for different initial conditions converge rapidly (after a time of the order $t\sim\epsilon$) to each other. The solution of (\ref{eq:reddyn}) almost coincides with those of the full dynamics, indicating that the latter can be approximated as dynamics on the manifold of distributions $\big(u_n^\text{fast}(x)\big)$. When $\epsilon$ increases, the solutions for different initial conditions differ more and more. This indicates that there are no longer autonomous dynamics in the variable $x$, and thus no well-defined specific growth rate.

\begin{table}
\caption{Parameter values used in the simulations}
\begin{center}
\begin{tabular}{ll}
\hline \\[-9pt]
floc growth rate $h_n(s)$
 & $h_n(s) = h(s) \, n^\alpha$ with $\alpha = \frac 23$ \\[3pt]
bacterium growth rate $h(s)$
 & $h(s) = \frac{0.2 \, s}{s+6}$ \\[3pt]
attachment coefficients $a_n$
 & $a_n = n^\alpha$ with $\alpha = \frac 23$ \\[3pt]
detachment coefficients $b_n$
 & $b_n = 0.1 \, n^\alpha$ with $\alpha = \frac 23$ \\[3pt]
dilution rate $D$
 & $D=0.04$ \\[3pt]
inflow substrate concentration $s_{\text{in}}$ \hspace{20pt}
 & $s_{\text{in}}=20$ \\[3pt]
\hline
\end{tabular}
\end{center}
\label{tab:compfullred}
\end{table}

We conclude that for larger values of $\epsilon$, the system cannot be described by a dynamical equation of the form (\ref{eq:reddyn}). Nevertheless, Fig.~\ref{fig:compfullred} shows that for all values of $\epsilon$, the different initial conditions lead to the same equilibrium. On the other hand, the equilibrium of (\ref{eq:reddyn}) different from the wash-out solution, i.e.
$x \neq 0$, satisfies $h(s,x)=D$. If we want the reduced dynamics to predict the correct equilibrium, the specific growth rate should satisfy this condition. In this way, we obtain a well-defined density-dependent growth rate, which we call the specific growth rate at equilibrium $h^{\text{equi}}(s,x)$.

\begin{figure}
\begin{center}
\includegraphics[width=6cm]{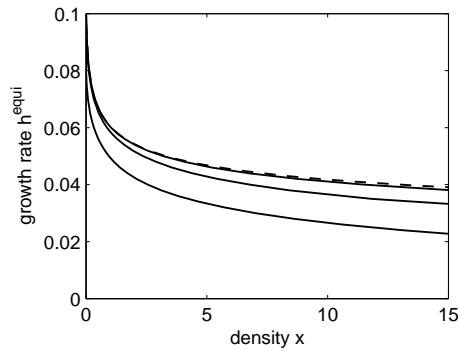}
\caption{The specific growth rate at equilibrium $h^{\text{equi}}(s,x)$ as a function of the biomass density $x$ for a fixed substrate density $s=6$. Parameter values of Tab.~\ref{tab:compfullred} were used. The curves in full line correspond to, from bottom to top, $\epsilon=1$, $\epsilon=0.1$ and $\epsilon=0.01$. The specific growth rate (\ref{eq:growthrate}) is shown in dashed line.}
\label{fig:growthequi}
\end{center}
\end{figure}

Fig.~\ref{fig:growthequi} plots the specific growth rate at equilibrium for different values of the parameter $\epsilon$. For small $\epsilon$, the specific growth rate at equilibrium coincides almost with the explicit formula (\ref{eq:growthrate}). As $\epsilon$ increases, the difference with (\ref{eq:growthrate}) becomes substantial. Moreover, the function $h^{\text{equi}}(s,x)$ no longer factorizes as a function of $s$ times a function of $x$. Note that all these curves are monotonically decreasing in the bacteria density $x$, as we proved explicitly for (\ref{eq:growthrate}).

\begin{figure}
\begin{center}
\raisebox{4.2cm}{(a)} \hspace{-.3cm}
\includegraphics[width=5.5cm]{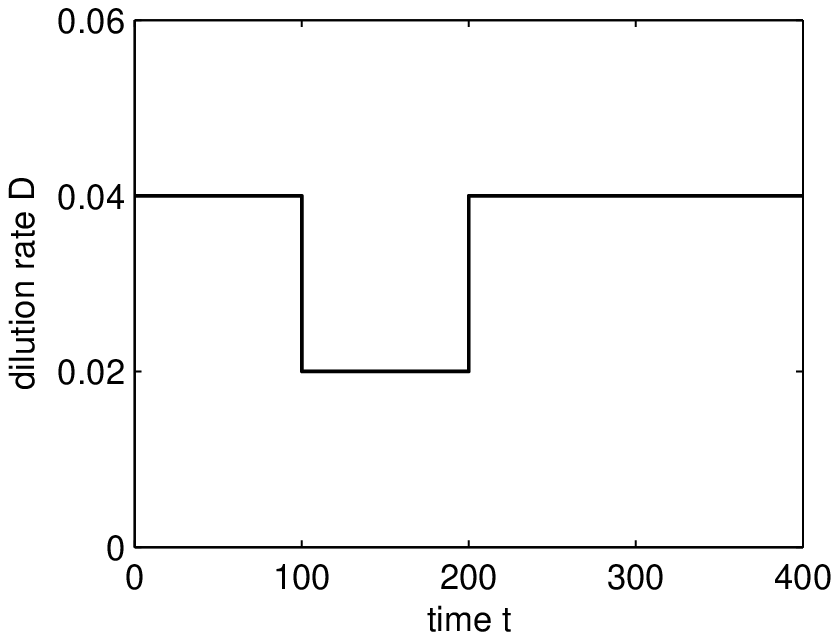}
\hspace{.5cm}
\raisebox{4.2cm}{(b)} \hspace{-.3cm}
\includegraphics[width=5.5cm]{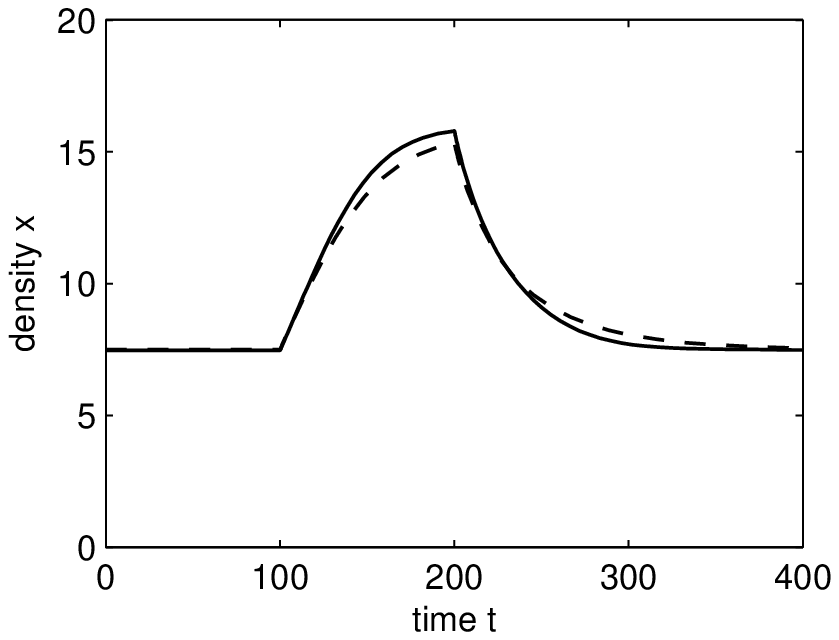}
\caption{Response of the full model (\ref{eq:fulldyn}) and the reduced model (\ref{eq:reddyn}) to a step in the dilution rate $D$. (a) Excitation in the dilution rate $D$. (b) Reaction of the two models. Parameter values of Tab.~\ref{tab:compfullred} were used. The full line corresponds to the system (\ref{eq:fulldyn}) with $\epsilon=1$. The dashed line corresponds to the system (\ref{eq:reddyn}) with specific growth rate $h^{\text{equi}}(s,x)$. For both simulations, the initial condition was taken as the equilibrium for dilution rate $D=0.04$.}
\label{fig:compfullred2}
\end{center}
\end{figure}

The reconstructed growth rates $h^{\text{equi}}(s,x)$ can now be used to integrate (\ref{eq:reddyn}), i.e.,
\begin{align*}
 \frac{\diff x}{\diff t} &= h^{\text{equi}}(s,x) x - D x \nonumber \\
 \frac{\diff s}{\diff t} &= -h^{\text{equi}}(s,x) x + D (s_\text{in}-s).
\end{align*}
By construction, this model will tend to the same equilibrium as the full model (\ref{eq:timescales}). To test how well it approximates the dynamics (\ref{eq:timescales}), we perturb the system out of equilibrium and look at the resulting dynamics. As shown in Fig.~\ref{fig:compfullred}, perturbations which
disturb too heavily the floc size distribution cannot be correctly modelled by an equation of the form (\ref{eq:reddyn}). We therefore apply a perturbation in the dilution rate $D$, which acts similarly on the different floc densities $u_n$. Fig.~\ref{fig:compfullred2} shows that the reduced model predicts with rather good precision the reaction of the full system to this perturbation.

\section{Conclusion}

In this paper we investigated how flocculation influences the bacterial growth dynamics in a bioreactor. In the context of the activated sludge process, this coupling of physico-chemical and biological phenomena is mostly relevant for the reaction tank. In particular, we studied the possibility of an effective model on the level of the biomass density, without explicitly taking flocculation into account.

Such an effective description is only possible when the flocculation dynamics are sufficiently fast compared to the other processes. In this case, the specific growth rate, which for isolated bacteria depends only on the substrate density, gains an additional dependence on the biomass density. It is interesting to note that such a density-dependent growth rate has recently been proposed as a mechanism to explain the coexistence of many bacterial species growing on a limited number of substrates. We will investigate the link between flocculation and species coexistence in a forthcoming contribution.

When the flocculation dynamics have time scales comparable to the bacterial growth, the details of the floc size distribution do affect the global system dynamics. In that case, dynamics autonomous in the biomass density do not exist, and the notion of specific growth rate is ill-defined. However, if the reactor evolves such that the floc size distribution remains equilibrated, it makes sense to define a specific growth rate at equilibrium. We showed in a simple example, that such a growth rate, which is again density-dependent, can yield an accurate description of the system dynamics.

\section*{Acknowledgements}

It is a pleasure to thank Roger Arditi, Denis Dochain, Nabil Mabrouk, Fr\'ed\'eric Mazenc, Alain Rapaport and Dimitri Vanpeteghem for valuable discussions.

\end{document}